\documentclass[prl, reprint, amsmath,amssymb,nofootinbib,floatfix, aps,longbibliography, preprintnumbers]{revtex4-1}
\usepackage{graphicx}
\usepackage{dcolumn}
\usepackage{bbold}
\usepackage{mathtools}
\usepackage[colorlinks=true,urlcolor=myblue,anchorcolor=black,citecolor=myblue,filecolor=black,linkcolor=black,menucolor=blue,linktocpage=true]{hyperref} 
\usepackage{booktabs}
\usepackage{colortbl}
\usepackage{collcell}
\usepackage{enumerate}
\usepackage{float}
\usepackage{verbatim}
\usepackage{multirow}
\usepackage{xparse}
\usepackage{slashed}
\usepackage{tabularx}
\usepackage{amsmath}
\usepackage{relsize}
\usepackage{braket}
\usepackage{xcolor}
\usepackage[compat=1.1.0]{tikz-feynman}
\usepackage{contour}

\newcommand{\virg}[1]{``#1''}

\definecolor{myred}{cmyk}{0,0.8,0.5,0.5}
\definecolor{myblue}{cmyk}{0.8, 0.4, 0, 0.2}
\definecolor{mygreen}{rgb}{0.27, 0.64, 0.48}
\definecolor{mygray}{gray}{.95}

\def\CR{{\mathsmaller{\mathrm{CR}}}}

\def\max{{\mathrm{max}}}
\def\min{{\mathrm{min}}}
\def\rel{{\mathrm{rel}}}
\def\ES{{\mathsmaller{\mathrm{ES}}}}
\def\DIS{{\mathsmaller{\mathrm{DIS}}}}
\def\BSM{{\mathsmaller{\mathrm{BSM}}}}

\begin{document}

\title{Relic Neutrino Background from Cosmic-Ray Reservoirs}

\author{Andrea Giovanni De Marchi}
\email{andreagiovanni.demarchi@unibo.it}
\affiliation{Dipartimento di Fisica e Astronomia, Università di Bologna, via Irnerio 46, 40126, Bologna, Italy;}
 \affiliation{INFN, Sezione di Bologna, viale Berti Pichat 6/2, 40127, Bologna, Italy}
 
 \author{Alessandro Granelli}
 \email{alessandro.granelli@unibo.it}
\affiliation{Dipartimento di Fisica e Astronomia, Università di Bologna, via Irnerio 46, 40126, Bologna, Italy;}
 \affiliation{INFN, Sezione di Bologna, viale Berti Pichat 6/2, 40127, Bologna, Italy}
 
\author{Jacopo Nava}
\email{jacopo.nava2@unibo.it}
\affiliation{Dipartimento di Fisica e Astronomia, Università di Bologna, via Irnerio 46, 40126, Bologna, Italy;}
 \affiliation{INFN, Sezione di Bologna, viale Berti Pichat 6/2, 40127, Bologna, Italy}
 
 \author{Filippo Sala}
 \email{f.sala@unibo.it}\thanks{FS is on leave from LPTHE, CNRS \& Sorbonne Universit\'{e}, Paris, France.}
\affiliation{Dipartimento di Fisica e Astronomia, Università di Bologna, via Irnerio 46, 40126, Bologna, Italy;}
 \affiliation{INFN, Sezione di Bologna, viale Berti Pichat 6/2, 40127, Bologna, Italy}

 \begin{abstract}

\noindent We compute the flux of relic neutrino background (R$\nu$B) up-scattered by ultra-high-energy (UHE) cosmic rays (CRs) in clusters that act as CR-reservoirs.
The long trapping times of UHECRs make this flux larger than that of R$\nu$B up-scattered by UHECRs on their way to Earth, which we also compute. We find that IceCube excludes R$\nu$B weighted overdensities larger than $10^{10}$ in clusters, and that PUEO, RNO-G, GRAND and IceCube-Gen2 will test values down to $10^{8}$. 
Our treatment incorporates the momentum transfer dependence of the neutrino-nucleus cross section, deep inelastic scattering, a mixed UHECR composition, and flavour information on the up-scattered R$\nu$B fluxes for both cases of neutrino mass spectrum with normal and inverted ordering, providing new handles to possibly disentangle the up-scattered R$\nu$B from cosmogenic neutrinos.

 \end{abstract}
\maketitle

\section{Introduction}
The relic neutrino background (R$\nu$B) is referred to as the \virg{Holy Grail} of neutrino physics. It is the only sub-component of dark matter that is predicted by the standard cosmological model ($\Lambda$CDM), with a present temperature and number density per-flavour (counting neutrinos and antineutrinos separately) of~\cite{Lesgourgues:2013sjj}
 \begin{equation}
 T_{\nu,0} \simeq 1.67\times 10^{-4}\,\text{eV},\quad 
 n_{\nu,0} \simeq 56\,\text{cm}^{-3}.
 \label{eq:standard_predictions}
 \end{equation}
Its detection would give new observational access to the earliest cosmological times ever probed. We have indirect evidence for it via early Universe observations~\cite{Planck:2018vyg,Steigman:2012ve}, but none of the existing techniques to detect the local R$\nu$B is expected to discover it in the foreseeable future, see~\cite{Bauer:2022lri} for a recent overview.
It has been furthermore pointed out that PTOLEMY~\cite{PTOLEMY:2019hkd,Apponi:2021hdu}, which aims at detecting the R$\nu$B by capture on tritium, is insensitive to it, if graphene is used 
to stock tritium, because of Heisenberg uncertainty~\cite{Cheipesh:2021fmg,PTOLEMY:2022ldz}.
Experiments aiming at detection of the local R$\nu$B then only test the case where its local density is much larger than the diffuse cosmological one $n_{\nu,0}$ by an overdensity factor $\eta_\nu^\text{Earth} > 1$. The strongest such limit has been set by the KATRIN experiment~\cite{Aker_2022} and reads $\eta_\nu^\text{Earth} < 1.3\times10^{11}$, in the same ballpark of limits from the R$\nu$B gravitational effects in the Solar System~\cite{Tsai:2022jnv}.
Given that gravitational clustering can induce at most $\eta_\nu^\text{Earth} \sim 10^2$~\cite{Ringwald:2004np}, these experiments then only test the beyond the Standard Model (BSM) scenarios, if any, that lead to those large local overdensities.

The main challenge to detect the R$\nu$B is its tiny energy, as a consequence of its low temperature $T_{\nu,0}$. 
A possible way-out consists in looking for consequences of the highest-energy scatterings that the R$\nu$B can undergo in the Universe, so to maximise their SM cross sections. To our knowledge, the first exploration along these lines was the computation of the R$\nu$B flux up-scattered by ultra-high-energy (UHE) cosmic rays (CRs), carried out by Hara and Sato in the 80's~\cite{Hara:1980ab,Hara:1980mz}, when much less than today was known about both neutrinos and UHECRs. 
This idea has been dormant for forty years, until the authors of~\cite{Ciscar-Monsalvatje:2024tvm} revived it and, from the non-observation of such an up-scattered R$\nu$B flux at IceCube, constrained overdensities in the ballpark of $10^{13}\,(10^{11})$ on scales of about $10\,\text{kpc}$ in the Milky Way (around the blazar TXS 0506+056).
While the study~\cite{Ciscar-Monsalvatje:2024tvm} is sufficient to set a rough limit, it has limitations, like assuming that all UHECRs are protons (which we know is not the case~\cite{PierreAuger:2022atd, TelescopeArray:2018bep}), as well as using an oversimplified SM cross section. These should be addressed in order to possibly hope to detect the up-scattered R$\nu$B flux and disentangle it from other neutrinos that could show up in a similar energy range, like cosmogenic ones.\footnote{
The R$\nu$B could also be indirectly tested via features that its scatterings induce in UHE neutrinos on their way to Earth~\cite{Fargion:1997ft,Weiler:1997sh}. IceCube observations constrain this way $\eta_\nu \lesssim 10^{11}\,(10^8)$ on scales of 10~kpc (of the 14~Mpc that separate us from NGC 1068)~\cite{Franklin:2024enc}.
Resonant dips in UHE cosmogenic neutrinos~\cite{Eberle:2004ua}, not observed so far~\cite{IceCube:2018fhm,PierreAuger:2019ens}, could at best test $\eta_\nu \sim 10^{11}$ on scales of the entire Universe~\cite{Brdar:2022kpu}, which are already ruled-out by not over-closing it.}

The R$\nu$B overdensities tested by the techniques above are not only considerably larger than those achievable via gravitational clustering~\cite{Ringwald:2004np}, but also violate the Pauli exclusion principle unless one introduces BSM physics that would cluster the R$\nu$B more than gravity, see~\cite{Bondarenko:2023ukx} for a systematic assessment.
To our knowledge, the largest overdensities achieved in a fully worked-out BSM model rely on a tiny long-range neutrino self-interaction~\cite{Smirnov:2022sfo}. They read $    \eta_{\nu}^{\BSM} \simeq
10^7 \,(m_\nu/0.1 \;\text{eV})^3$, where $m_\nu$ is the neutrino mass scale, and extend up to a volume of the size of a galaxy cluster, motivating to test large overdensities in these environments.

In this manuscript, 
we propose to look for the R$\nu$B up-scattered by UHECRs in clusters that act as CR-reservoirs~\cite{Volk:1994zz, Berezinsky:1996wx, Murase:2008yt, Kotera:2009ms, Murase:2012rd, Fang:2017zjf, Condorelli:2023xkx} and calculate the associated fluxes. This idea benefits from the long trapping times of CRs in these environments, and from the knowledge on them that is available today and will increase with upcoming observations in the near future. 
Additionally, we improve over~\cite{Ciscar-Monsalvatje:2024tvm} in the calculation of UHECR-up-scattered R$\nu$B fluxes in a number of ways, which we will show to be quantitatively important.\\

\section{Up-scattering the relic neutrino background with cosmic rays}
The flux of accelerated relic neutrinos per unit energy $E_\nu$, from the up-scattering environments of interest, can be written as
   \begin{equation}
   \label{eq:dPhinudE}
    \frac{d\Phi_\nu}{dE_\nu} = D_\text{eff}\,n_{\nu,0}\, \bar{\eta}_\nu\int_{E_\CR^\min(E_\nu)}^{E_\CR^\max}dE_\CR\frac{d\Phi_\CR}{dE_\CR} \frac{d\sigma_{\nu \CR}}{dE_\nu},
\end{equation}
 where the effective distance $D_\text{eff}$ contains information on the spatial distribution of relic neutrinos and CRs; 
$\bar{\eta}_\nu\equiv (1/V)\int_V d^3\vec{r}\ \eta_\nu(\vec{r})$ is the R$\nu$B overdensity averaged over the volume $V$, with $\eta_\nu(\vec{r}) = n_\nu(\vec{r})/n_{\nu,0}$ and $n_\nu(\vec{r})$ being the neutrino number density at position $\vec{r}$;  $E_\CR^\min(E_\nu)$ is the greatest value between the minimal available CR energy and the lowest energy required by the kinematics of the scattering; $E_\CR^\max$ is the largest CR energy in their flux; $d\Phi_\CR/dE_\CR$ is the CR flux per unit of CR energy $E_\CR$; $d\sigma_{\nu\CR}/dE_\nu$ is the differential cross section for the $\nu$-CR scattering. The dependence on the up-scattering environment, in this letter either the Milky Way or CR-reservoirs, is contained in $D_\text{eff}$ and $d\Phi_\CR/dE_\CR$. The total neutrino flux is understood to be the sum of Eq.~\eqref{eq:dPhinudE} over all neutrinos and nuclear species composing CRs.\\

\subsection{Neutrino-nucleus scattering at UHEs}
We consider the up-scattering of a relic (anti)neutrino $\nu$ ($\bar{\nu}$) by a cosmic nucleus $\mathcal{N}$ via pure SM neutral current (NC) interaction.
At the $\nu$-$\mathcal{N}$ exchanged energies that we are interested in, the scattering is described by that on the nucleons $N=p,n$ as
  \begin{equation}
  \frac{d\sigma_{\nu\mathcal{N}}}{dE_\nu}\left(E_\mathcal{N}\right) \simeq
  \frac{A_\mathcal{N}}{2}\left[\frac{d\sigma_{\nu p}}{dE_\nu}\left(\frac{E_\mathcal{N}}{A_\mathcal{N}}\right) + \frac{d\sigma_{\nu n} }{dE_\nu}\left(\frac{E_\mathcal{N}}{A_\mathcal{N}}\right)\right],
  \end{equation}
where we have focused for simplicity on isoscalar nuclei with equal number $A_\mathcal{N}/2$ of protons and neutrons, each carrying a fraction $1/A_\mathcal{N}$ of the nucleus's energy $E_\mathcal{N}$,
and summed over elastic scattering (ES) and deep inelastic scattering (DIS) contributions: $d\sigma_{\nu N}/dE_\nu = d\sigma_{\nu N}^\ES/dE_\nu + d\sigma_{\nu N}^\DIS/dE_\nu$. 
The ES part, summed over $\nu$ and $\bar{\nu}$, reads~\cite{Giunti:2007ry, Formaggio:2012cpf}
\begin{equation}\label{eq:ES_xsec}
    \frac{d\sigma_{\nu N}^{\ES}}{dE_\nu} = \frac{2G_F^2 m_\nu m_N^4}{\pi(s-m_N^2)^2}\left[A_N(Q^2) + C_N(Q^2)\frac{(s-u)^2}{m_N^4}\right],
\end{equation}
where $G_F$ is the Fermi constant; $s=  2m_\nu E_N + m_N^2 + m_\nu^2$, $Q^2=2m_\nu (E_\nu-m_\nu)$ 
is the momentum transfer squared, $u = 2m_\nu^2 + 2m_N^2 -s +Q^2$;
$m_{N(\nu)}$ is the nucleon (neutrino) mass and $E_{N(\nu)}$ is the energy of the incoming nucleon (outgoing neutrino) in the frame in which the initial neutrino is at rest.
The functions $A_N$ and $C_N$ are given in Appendix~\ref{app:FormFacts_ES} (see, e.g.,~\cite{Giunti:2007ry, Formaggio:2012cpf}) and strongly suppress ES for $Q^2 \gtrsim m_N^2 \approx\text{GeV}^2$. 
The DIS, in which the initial neutrino interacts directly with the quark constituents of the nucleons, takes over for $E_N \gtrsim 10^{10-11}$~GeV, given $Q^2 \lesssim s$ and $m_\nu \approx 0.1\,\text{eV}$, 
thus being necessary to describe the highest-energy part of the up-scattered R$\nu$B flux. For the NC $\nu$-$N$ DIS cross section we adopt
\begin{equation}\label{eq:DIS_xsec}
\begin{split}
    \frac{d\sigma^{\DIS}_{\nu N}}{dE_\nu} \simeq & \sum_{a=q,\bar{q}}\frac{G_F^2[(g_V^a)^2 + (g_A^a)^2]}{2\pi E_N}\\
    &\times \int_{y_\min}^1\frac{dy}{y^2}\,\frac{Q^2f_a^N(x,Q^2)}{[1+Q^2/M_Z^2]^2}\,
   g(y, Q^2, m_N),
    \end{split}
\end{equation}
where $g(y, Q^2, mN) \equiv  (y^2-2y+2-2m_N^2x^2 y^2/Q^2)$, $M_Z$ is the $Z$ boson mass, $y$ is the inelasticity parameter satisfying $y_\min = (E_\nu-m_\nu)/E_N \lesssim y \leq 1$, $x = (E_\nu-m_\nu)/(E_N y)$ is the Bjorken scaling variable and $f_a^N(x, Q^2)$ is the parton distribution function (PDF) for the quark $a$ having NC vector and axial coupling $g_V^a$ and $g_A^a$, $a = u, d, s, c, b$ (we neglect any contribution from the top quark). We evaluate the PDFs with the Python package \texttt{parton} using the \virg{CT10} PDF set~\cite{Lai:2010vv} (see also \href{https://lhapdf.hepforge.org/}{this website} for more details). 
For simplicity, we do not take the coherent $\nu$-$\mathcal{N}$ scattering into account as it could contribute only at energy scales smaller than those of interest to our study~\cite{Freedman:1977xn, Formaggio:2012cpf}, and neglect the contribution to $d\sigma_{\nu\mathcal{N}}/dE_\nu$ from  hadronic resonances~\cite{Feynman:1971wr, Rein:1980wg}, relevant for $Q^2\approx\text{GeV}^2$, so that in the considered range our results are conservative. Details on the derivation of Eq.~\eqref{eq:DIS_xsec} are given in Appendix~\ref{app:DIS}.\\

\subsection{CRs up-scattering R$\nu$B \textit{en route} to Earth}
As they travel towards the Earth, UHECRs up-scatter the R$\nu$B with the largest cross sections among all CRs, and induce a flux described by Eq.~(\ref{eq:dPhinudE}). While the bulk of CRs with energy much below the EeV are believed to originate within the Milky Way (MW)~\cite{Gabici:2019jvz, BeckerTjus:2020xzg} and to be mostly protons below the PeV~\cite{Thoudam:2016syr}, UHECRs above the EeV scale are today understood as having an extragalactic origin, see e.g.~\cite{AlvesBatista:2019tlv,PierreAuger:2020fbi, Coleman:2022abf}, and a mixed composition with heavier nuclei dominating over protons, see e.g.~\cite{HiRes:2004hvo, Aloisio:2013hya, Abbasi:2014sfa, TelescopeArray:2018bep, TelescopeArray:2020bfv,PierreAuger:2022atd, PierreAuger:2023htc, Lv:2024wrs}. We therefore employ the CR spectrum $d\Phi_\CR/dE_\CR$ from~\cite{PierreAuger:2022atd} and, by considering up-scatterings in the MW with the line-of-sight integral $D_\text{eff} \approx 10$~kpc as in~\cite{Ciscar-Monsalvatje:2024tvm} (see e.g.~\cite{Guo:2020oum, Xia:2021vbz, CDEX:2022fig} for studies of the uncertainties on $D_\text{eff}$, where the R$\nu$B case is analogous to relativistic DM fluxes and cored profiles like Isothermal), we obtain the flux reported as a dot-dashed line in Fig.~\ref{fig:Boosted_nu_flux} (see further). Our calculation improves over the analogous one~\cite{Ciscar-Monsalvatje:2024tvm} by going beyond the proton-only composition of CRs and by implementing the momentum transfer dependence and DIS in the cross section. It results in a significantly different flux with respect to~\cite{Ciscar-Monsalvatje:2024tvm}, see Appendix~\ref{app:comparison} for details of the comparison.

Since tentative sources of UHECRs are located at $1\sim 100\,\text{Mpc}$ from Earth~\cite{Wibig:2007pf, Taylor:2011ta, Mollerach:2019wne, Lang:2020wrr, Plotko:2022urd}, $D_\text{eff}$ can be larger than 10~kpc by orders of magnitude, resulting in a sensitivity to $\bar{\eta}_\nu \sim 10^{10-11}$. Still, these enormous $\bar{\eta}_\nu$ over such long distances are hard to justify without spoiling the large-scale homogeneity of the Universe, motivating us to look for environments where $D_\text{eff}$ can be sizeable while keeping neutrino overdensities localised on smaller scales.\\

\subsection{Boosted R$\nu$B from CR-reservoirs}
A possible explanation for the extragalactic origin of UHECRs is that they are produced inside galaxy clusters, in which they reside up to cosmological times before escaping
and contributing to the UHECR flux on Earth 
\cite{Volk:1994zz, Berezinsky:1996wx, Murase:2008yt, Kotera:2009ms, Murase:2012rd, Fang:2017zjf, Condorelli:2023xkx} (see also~\cite{Brunetti:2014gsa} for a review). 
The distance they travel inside these gigantic CR-reservoirs would be roughly $c\tau_\text{esc} \simeq 0.3\,\text{Gpc}\, (\tau_\text{esc}/1\,\text{Gyr})$, where $\tau_\text{esc}$ is the time CRs spend in the cluster before their release. 
The effective distance entering Eq.~(\ref{eq:dPhinudE}) can be written as $D_\text{eff} = \mathcal{B} c \tau_\text{esc}$, having defined the spatial boost factor 
\begin{equation}
    \mathcal{B} \equiv \int_V d^3\vec{r} f_\CR(x)\,\delta_\nu(\vec{r})\,,
    \label{eq:boost}
\end{equation}
 with  $\delta_\nu(\vec{r})\equiv \eta_\nu(\vec{r})/\bar{\eta}_\nu$
and $f_\CR(\vec{r})$ the spatial profile 
of the CR flux inside a CR-reservoir of volume $V$, normalised such that $\int_V d^3\vec{r}f_\CR(\vec{r})=1$.
One has $\mathcal{B}=1$ for a homogeneous R$\nu$B, $\delta_\nu(\vec{r})=1$, while $\mathcal{B} > 1$ if, e.g., the number densities of both UHECRs and neutrinos are peaked at small radii.
In what follows, we fix $\mathcal{B} =1 $ and $\tau_\text{esc}\simeq 2\,\text{Gyr}$~\cite{Fang:2017zjf}, neglecting any dependence on the CR energy that $\tau_\text{esc}$ (and $\mathcal{B}$) may have. We checked that implementing the $\tau_{\text{esc}}$ energy-dependence as in \cite{Condorelli:2023xkx} would only 
slightly (enhance) suppress the (low-energy) high-energy tail of the boosted R$\nu$B spectrum, without altering significantly our final results.

\begin{figure}[t]
    \centering
    \includegraphics[width = 0.48\textwidth]{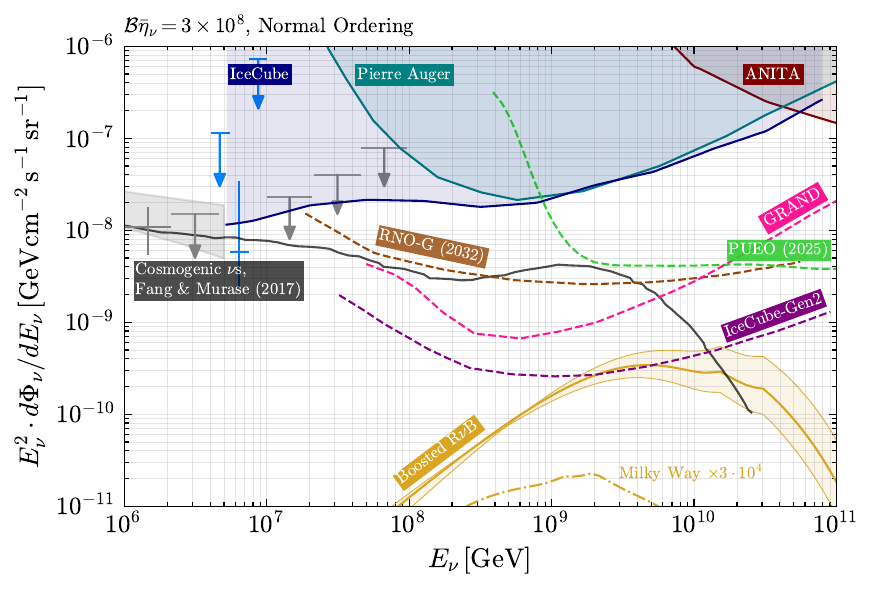}
    \includegraphics[width = 0.48\textwidth]{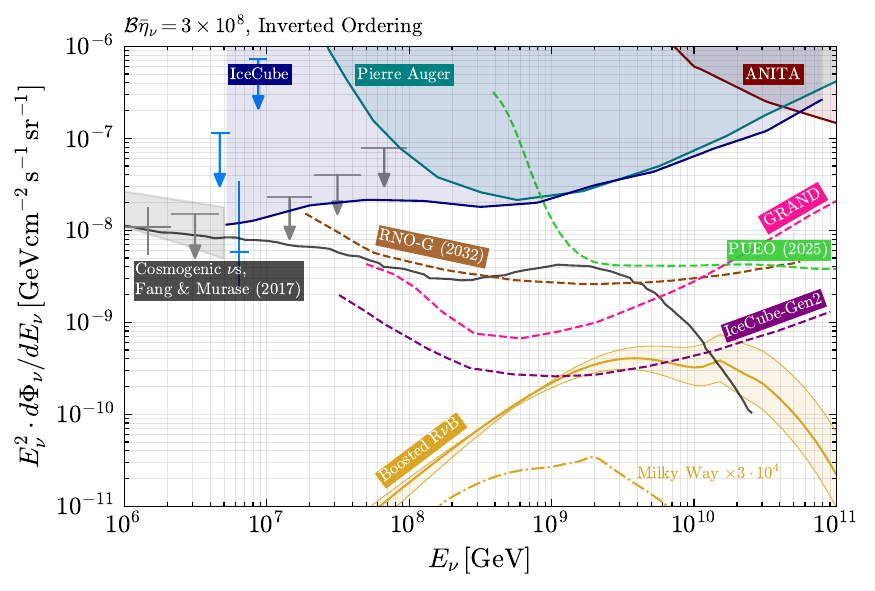}
    \caption{All-flavour R$\nu$B fluxes on Earth as up-scattered by UHECRs inside
    galaxy clusters acting as 
    CR-reservoirs (continuous gold lines) or from UHECRs in the MW (dot-dashed gold line, multiplied by $3\times 10^4$). The yellow shaded region is obtained for $2\leq \alpha\leq 2.5$ (central thick line, $\alpha = 2.3)$, 
    a weighted R$\nu$B overdensity
    $\mathcal{B}\bar{\eta}_\nu = 3\times10^8$, a neutrino mass spectrum with NO (IO) in the upper (lower) panel and $\sum_i m_i = 0.113 \,(0.145)\,\text{eV}$~\cite{DESI:2024mwx}.
    In grey and azure are the astrophysical neutrinos events observed at IceCube~\cite{IceCube:2020acn, IceCube:2021rpz} (see also~\cite{IceCube:2020wum, Abbasi:2024jro}, the KM3NeT ARCA detector will have a similar sensitivity~\cite{KM3NeT:2024uhg}). 
    Fluxes of UHE neutrinos lying in the upper shaded regions are excluded at $90\%$ C.L.~by the null-detection at IceCube~\cite{IceCube:2018fhm}, Pierre Auger Observatory~\cite{PierreAuger:2019ens} and ANITA~\cite{ANITA:2019wyx}.
    Sensitivities of PUEO (2025)~\cite{PUEO:2020bnn}, RNO-G (2032, first stations already taking data)~\cite{RNO-G:2023ibt}, IceCube-Gen2~\cite{IceCube-Gen2:2021rkf} (planned) and GRAND~\cite{GRAND:2018iaj} (proposed) are depicted as dashed lines. The black line displays the cosmogenic neutrino flux from~\cite{Fang:2017zjf}.}
    \label{fig:Boosted_nu_flux}
\end{figure}

We model UHECRs in cluster reservoirs following~\cite{Fang:2017zjf}. For each nucleus $\mathcal{N}$, we write
\begin{equation}\label{eq:CRflux_res}
\frac{d\Phi_{\mathcal{N}}}{dE_\mathcal{N}} = K_\mathcal{N}\left(\frac{E^\max_\mathcal{N}}{E_{\mathcal{N}}}\right)^\alpha e^{-E_{\mathcal{N}}/E^\max_\mathcal{N}},
\end{equation}
where $2\leq \alpha \leq 2.5$ and $E^\max_\mathcal{N}/Z_{\mathcal{N}} \simeq 7.69\times 10^{10}\text{GeV}$, $Z_\mathcal{N}$ being the atomic number of $\mathcal{N}$. We checked that varying $E^\max_\mathcal{N}$ within a factor of up to five from our benchmark choice has a comparable impact, on the detectability of the R$\nu$B flux, to varying $2\leq \alpha \leq 2.5$. We consider the relative nuclear abundances as given in~\cite{Fang:2017zjf} and fix the normalisation factors $K_\mathcal{N}$ by requiring that the total luminosity emitted from the entire population of CR-reservoirs matches with the one observed at Earth.
Concentrating on the energy range $ E_\CR^\text{ankle} = 5~\text{EeV} \leq E_\CR \leq 200\,\text{EeV}$, approximating the observed extragalactic CR spectrum as a single power-law $d\Phi_\CR/dE_\CR \simeq (d\Phi_\CR/dE_\CR)_\text{ankle}\, (E_\CR^\text{ankle}/E_\CR)^{2.5}$, with $(d\Phi/dE_\CR)_\text{ankle} \simeq 10^{-27}\,\text{GeV}^{-1}\text{cm}^{-2}\text{s}^{-1}\text{sr}^{-1}$~\cite{PierreAuger:2022atd}, and taking as a benchmark value $\alpha = 2.3$~\cite{Fang:2017zjf}, we find $K_\text{H(He)} \simeq 9.44\times 10^{-31}\,(9.52\times 10^{-32})\,\text{GeV}^{-1}\text{cm}^{-2}\text{s}^{-1}\text{sr}^{-1}$ for ${}^1\text{H}$ (${}^4\text{He}$), while heavier nuclei are sub-dominant. We give more details on the normalisation procedure in Appendix~\ref{app:normalisation}.

The resulting total flux on Earth of relic neutrinos up-scattered in CR-reservoirs is shown in Fig.~\ref{fig:Boosted_nu_flux}. We compare it with current observations, limits and future sensitivities, and with the flux of cosmogenic neutrinos arising as secondary products of UHECRs interactions inside reservoirs as computed in~\cite{Fang:2017zjf}, because that could constitute a background to our signal.
The total flux is obtained by summing over all neutrino mass eigenstates $\nu_i$ having non-zero masses $m_i$, $i=1,2,3$. The upper and lower panels are respectively for a light neutrino mass spectrum with normal ordering (NO) $m_1<m_2<m_3$, and inverted ordering (IO) $m_3<m_1<m_2$. For NO (IO), the squared mass differences are fixed to $\Delta m^2_{21}\equiv m_2^2-m_1^2 = 7.42\times 10^{-5}\,\text{eV}^2$ and $\Delta m^2_{31(23)}\equiv m_{3(2)}^2-m_{1(3)}^2 = 2.507 \,(2.486)\,\times 10^{-3}\,\text{eV}^2$~\cite{Esteban:2020cvm}, the sum of neutrino masses saturates the cosmological limit $\sum_i m_i = 0.113\,(0.145)\,\text{eV}$~\cite{DESI:2024mwx}, and the weighted overdensity is set to $\mathcal{B}\bar{\eta}_\nu = 3\times 10^8$ for concreteness.

We further note that R$\nu$B-UHECRs scatterings could in principle also alter the  measured spectrum of UHECRs. While a detailed study goes beyond the purpose of this work, here we simply estimate the mean free path of UHECRs in CR-reservoirs, due to R$\nu$B scatterings as $\lambda \sim 1/(\sigma_{\nu\CR}\bar{\eta}_\nu n_{\nu,0})$, finding that the R$\nu$B has a smaller impact on UHECRs than the CMB as long as $\bar{\eta}_\nu \lesssim 10^9$.\\

\section{Limits and sensitivities on relic neutrino overdensities}
\begin{figure}[t]
\centering
    \includegraphics[width = .47\textwidth]{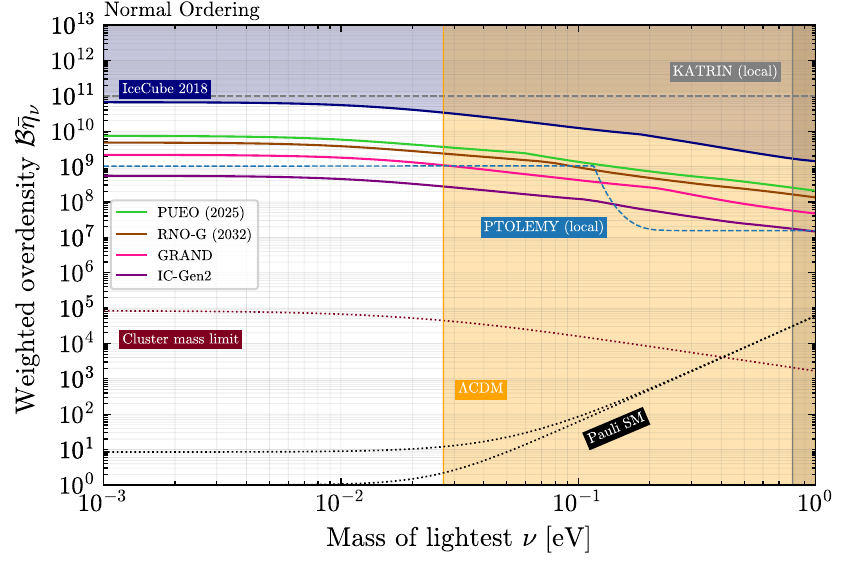}
    \includegraphics[width = 0.47\textwidth]{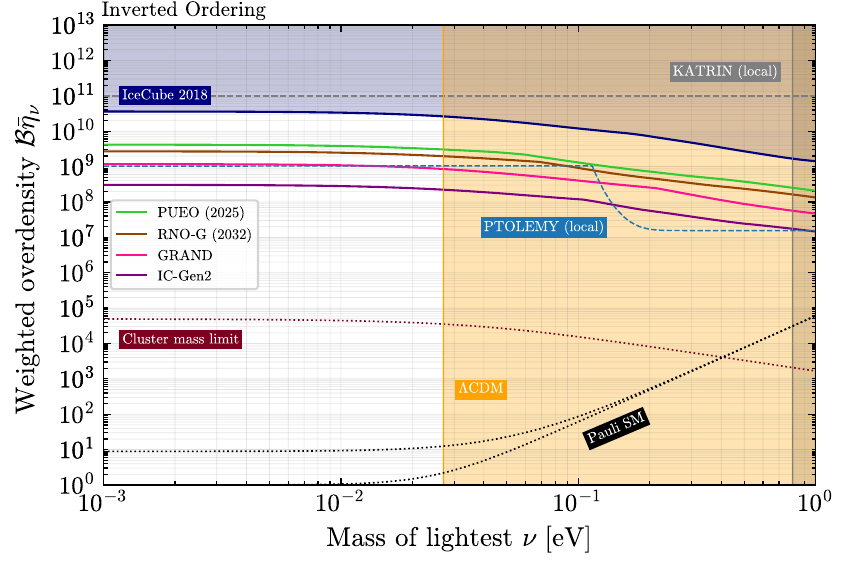}
\caption{Limits and sensitivities of $\mathcal{B}\bar{\eta}_\nu$, versus the lightest
neutrino mass. Assuming the reference slope $\alpha = 2.3$ in the CR spectrum in CR-reservoirs, and a neutrino mass spectrum with NO (IO) in the upper (lower) panel, we show the $90\%$ C.L.~constraint set by the null-detection at IceCube~\cite{IceCube:2018fhm} (blue shaded) and the sensitivities at PUEO (2025)~\cite{PUEO:2020bnn}, RNO-G (2032)~\cite{RNO-G:2023ibt}, GRAND~\cite{GRAND:2018iaj} and IceCube-Gen2~\cite{IceCube-Gen2:2021rkf} (solid lines, top to bottom).
Also shown are: the KATRIN limits on neutrino masses (solid), $m_\nu < 0.8 \,\text{eV}$ at $90\%$ C.L.~\cite{KATRIN:2021uub}; the KATRIN limit on local neutrino overdensities (dashed gray), $\eta_\nu^\text{Earth} < 9.7\times 10^{10}$~\cite{Aker_2022}; the limit on the sum of neutrino masses from DESI, $\sum_i  m_i < 0.113\,(0.145) \,\text{eV}$ at $95\%$ C.L.~assuming $\Lambda$CDM \cite{DESI:2024mwx} (orange shaded); the maximum $\mathcal{B}\bar{\eta}_\nu$ allowed by the Pauli exclusion principle (black dotted, $\mathcal{B} = 1$), respectively for the heaviest (upper line) and lightest (lower line) neutrino; the cluster mass limit (red dotted); the PTOLEMY sensitivity (dashed cyan) on the local overdensity \cite{PTOLEMY:2022ldz}. See the text for further details.}
 \label{fig:etanu_cons}
\end{figure}

In Fig.~\ref{fig:etanu_cons} we show the limits and sensitivities on the combination $\mathcal{B}\bar{\eta}_\nu$, against the lightest neutrino mass assuming NO (IO) in the upper (lower) panel.
We derive them by imposing that our flux line touches the relevant limit/sensitivity curve. These limits and sensitivities are shown as solid curves in Fig.~\ref{fig:etanu_cons} with different colours, depending on the experiment (see the legend in the plot and the caption for more details).

Our limits and sensitivities are compared with the Pauli exclusion principle constraining the number density of gravitationally-bound neutrinos, when no BSM clustering effect is assumed. We compute the local maximum overdensity as prescribed in~\cite{Ringwald:2004np} for a galaxy cluster of mass $M=5\times10^{15} M_\odot$, assuming a Navarro-Frenk-White (NFW)~\cite{Navarro_1997} profile for the surrounding dark matter halo, and then average over the cluster volume. The
results of this procedure are shown in dotted black in Fig.~\ref{fig:etanu_cons}.

We also evaluate the overdensity $\bar{\eta}_\nu$ for which the R$\nu$B mass equals that of the entire galaxy cluster. We consider a NFW dark matter density profile whose virial radius scales as $R_\text{vir} \sim M^{1/3}$, such that the average mass density in clusters $\rho_\text{cluster}$ is independent of the cluster mass. In particular, $\rho_\text{cluster} \approx 200\rho_c$, with $\rho_c \simeq 1.05\times 10^{-5}h^{-2}\,\text{GeV}\,\text{cm}^{-3}$ the critical density of the Universe and $h\simeq 0.674$ the dimensionless Hubble parameter \cite{Workman:2022ynf}.
We then impose a rough cluster mass limit by requiring $\bar{\eta}_\nu n_{\nu,0}\sum_i m_i \leq \rho_\text{cluster}$, taking equal averaged overdensity for each neutrino species. The corresponding limit on the weighted overdensity $\mathcal{B}\bar{\eta}_\nu$ can be relaxed for non-uniform neutrino and CR spatial distributions. This
is shown in dotted red in Fig.~\ref{fig:etanu_cons}.

Finally, we estimate the sensitivity of the proposed PTOLEMY experiment~\cite{PTOLEMY:2022ldz} on local neutrino overdensities, 
considering the configuration where the final ${}^3\text{He}^+$ is in the bound ground state. Depending on the neutrino mass compared to the experimental resolution $\Delta$ (we use $\Delta = 0.05$ eV as reference), the R$\nu$B absorption peak can be well-separated from the continuous $\beta$-decay spectrum or hidden under it. Accordingly,
we require at least $N_\text{peak} = 10\;\text{events/yr}$ to call a detection in the first situation, or $N_\text{peak}\gtrsim 3\sqrt{N_\text{bkg}}$, with $N_\text{bkg}$ the number of $\beta$-decay events/yr in the same energy range in the second. The PTOLEMY sensitivity is shown in dashed cyan
in Fig.~\ref{fig:etanu_cons}. 

For IO and in the hierarchical limit with $m_3 \ll m_1 \lesssim m_2$, there are
two heavy and one light neutrino implying a flux that is
larger by a factor $\sim 2$ with respect to the results in the
NO case. Correspondingly, the limits and sensitivities on $\mathcal{B}\bar{\eta}_\nu$ are slightly improved compared to the case of NO.
We note, however, that also the cluster mass limit becomes
more stringent by an equal amount, for the same
reason. In the degenerate limit $m_1 \simeq m_2 \simeq m_3$, the IO case is practically identical to the scenario with NO.\\

\section{Flavour composition of up-scattered relic neutrinos}
Neutrinos produced in astrophysical environments typically exhibit precise flavour composition at the source, but neutrino oscillations over astronomical distances tend to homogenise any flavour disparity~\cite{Palladino:2015vna, Bustamante:2015waa}.

Instead, the boosted R$\nu$B exhibits a peculiar flavour composition. At first, the R$\nu$B is evenly distributed among the mass eigenstates, which directly take part in the NC scatterings with UHECRs. The mass eigenstates $\nu_i$ then propagate freely  and their fluxes $d\Phi_i/dE_\nu$ are preserved. At detection, the probability of observing a massive neutrino $\nu_i$ in a  flavour $\ell=e,\,\mu,\,\tau$ is $\mathcal{P}_{\ell i} = \left|U_{\ell i}\right|^2$, with $U_{\ell i}$ being the entries of the Pontecorvo-Maki-Nakagawa-Sakata (PMNS) neutrino mixing matrix~\cite{Pontecorvo:1957cp, Pontecorvo:1957qd, Maki:1962mu}. Then, the flux of boosted neutrinos with flavour $\ell$ is given by $ d\Phi_\ell/dE_\nu = \sum_i \left|U_{\ell i}\right|^2 d\Phi_i/dE_ \nu$. Clearly, the relative flux in each flavour depends on the neutrino mixing parameters through the PMNS matrix and neutrino masses via the differential flux.

\begin{figure}
    \centering
    \includegraphics[width = 0.45\textwidth]{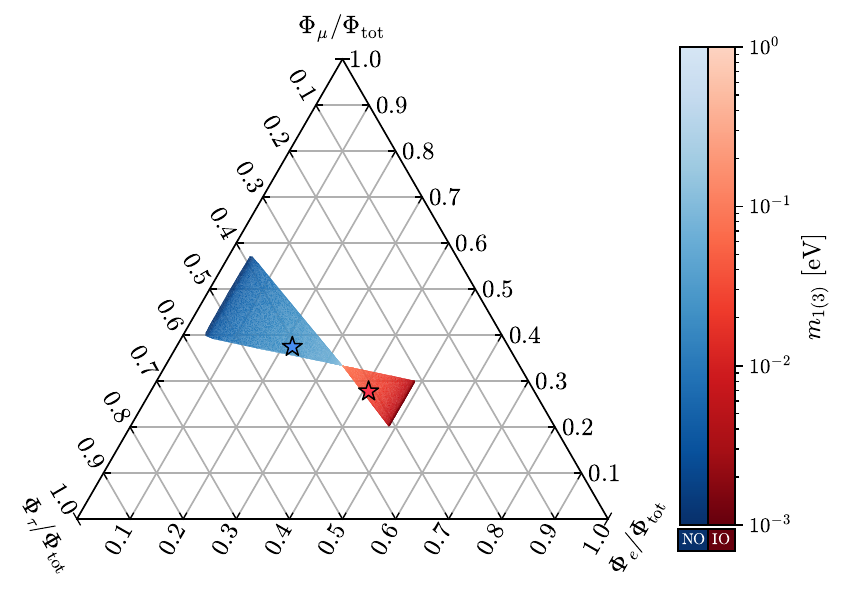}
    \caption{Flavour composition of the boosted R$\nu$B. The blue (red) area is obtained by varying the PMNS matrix entries and squared neutrino mass splittings within the 3$\sigma$ allowed range~\cite{Esteban:2020cvm}, assuming NO (IO). Darker regions correspond to lighter $m_{1(3)}$. The stars mark the flavour composition of the boosted R$\nu$B for the best-fit values of the neutrino oscillation data and $m_{1(3)}$ saturating the cosmological bounds $\sum_i m_i = 0.113\,(0.145)\,\text{eV}$~\cite{DESI:2024mwx} (the flavour composition depends only very slightly on the slope of the CR spectrum).}
    \label{fig:Ternary_plot}
\end{figure}

As the fluxes of the different mass eigenstates have distinct energy dependence, the flavour composition of the boosted R$\nu$B flux depends on the energy as well. It is nevertheless practical to calculate an integrated flavour ratio $\Phi_\ell/\Phi_{\text{tot}} \equiv \sum_i |U_{\ell i}|^2\Phi_i/\sum_i \Phi_i$. The predicted integrated flavour composition is shown in Fig.~\ref{fig:Ternary_plot}. In the plot, the parameters of the PMNS matrix are varied within the $3\sigma$ ranges allowed by the \texttt{NuFit 5.2} global analysis of neutrino oscillation data~\cite{nufit,
Esteban:2020cvm}, while $m_{1(3)}$ in the range $10^{-3}\leq m_{1(3)}/\text{eV}\leq 1$, for NO (IO). As $m_{1(3)}$ increases, the mass eigenstates become nearly degenerate, implying $\Phi_1\simeq \Phi_2\simeq \Phi_3$ and an unflavoured composition $\Phi_e:\Phi_\mu:\Phi_\tau = 1:1:1$, due to the unitarity of the PMNS matrix. By decreasing $m_{1(3)}$, the mass spectrum becomes hierarchical with $m_1 \lesssim m_2 \ll m_3$ ($m_3 \ll m_1 \lesssim m_2$) with the (two) heaviest neutrino(s) dominating the total flux. In this situation, because of the structure of the PMNS matrix, the flavour flux composition is approximately $0:1:1$ ($2:1:1$) for NO (IO). This could help UHE neutrino observatories discriminate between a boosted R$\nu$B signal and other kinds of astrophysical neutrino fluxes~\cite{IceCube:2015rro, Palladino:2019pid, GRAND:2018iaj, PUEO:2020bnn, IceCube-Gen2:2020qha, Coleman:2024scd}.\\

\section{conclusions}
We computed the flux of relic neutrinos up-scattered by UHECRs via SM NC interactions, taking into account the full $Q^2$-dependence and DIS in the cross section, and the mixed CR composition. Our results are shown in Fig.~\ref{fig:Boosted_nu_flux} for up-scatterings in the MW by UHECRs travelling towards Earth and in galaxy clusters acting as CR-reservoirs. These objects constitute an ideal environment because of the long trapping times of CRs inside them, and indeed lead to the largest R$\nu$B up-scattered fluxes. Our calculation is conservative, having not included the hadron-resonances contributions to $\nu$-UHECR scatterings, nor the secondary neutrinos produced by their SM charged-current interactions.

We find that IceCube~\cite{IceCube:2018fhm} excludes $\mathcal{B}\bar{\eta}_\nu \gtrsim 10^{10}$ in galaxy clusters, where we weighted the average overdensity $\bar{\eta}_\nu$ by a spatial boost factor $\mathcal{B}$, and that future telescopes~\cite{PUEO:2020bnn,RNO-G:2023ibt,GRAND:2018iaj,IceCube-Gen2:2021rkf} could possibly detect the boosted R$\nu$B for $\mathcal{B} \bar{\eta}_\nu \gtrsim 10^8$, see Fig.~\ref{fig:etanu_cons}. These large overdensities require a BSM origin, as could be obtained on the scales of galaxy clusters via, e.g.,  
long-range interactions.

To distinguish a boosted R$\nu$B signal from other UHE$\nu$ fluxes, such as cosmogenic neutrinos, we propose to rely on i) the shape of the energy spectrum, DIS being crucial in determining the one of the up-scattered R$\nu$B, and ii) 
the flavour composition, the specific one of the up-scattered R$\nu$B being displayed in Fig.~\ref{fig:Ternary_plot}.

Our study motivates further UHE$\nu$ searches at telescopes, a direct implementation of R$\nu$B-UHECR scatterings in the modelling of CR-reservoirs and sources, and research on how to obtain the sizeable neutrino overdensities required for a potential detection.\\

\begin{acknowledgments}

\section*{Acknowledgements}
We thank Angelo Esposito for precious elucidations on the PTOLEMY experimental set-up, Ke Fang for clarifications on~\cite{Fang:2017zjf}, and Silvia Pascoli for useful discussions.
 We acknowledge the use of computational resources from the parallel computing cluster of the Open Physics Hub (\href{https://site.unibo.it/openphysicshub/en}{https://site.unibo.it/openphysicshub/en}) at the Physics and Astronomy Department in Bologna. 
 J.N. acknowledges hospitality from the Fermilab Theoretical Physics Department.
 This work was supported in part by the European Union's Horizon research and innovation programme under the Marie Skłodowska-Curie grant agreements No.~860881-HIDDeN and No.~101086085-ASYMMETRY, by COST (European Cooperation in Science and Technology) via the COST Action COSMIC WISPers CA21106, and by the Italian INFN program on Theoretical Astroparticle Physics.

\end{acknowledgments}



\appendix
\section{Form Factors of the $\nu$-$N$ Neutral Current Elastic Scattering Cross Section}
\label{app:FormFacts_ES}
A detailed derivation of the $\nu$-$N$ elastic scattering cross section can be found in, e.g.,~\cite{Giunti:2007ry} (see also~\cite{Formaggio:2012cpf}). Here, for completeness, we only report the form of the factors $A_N$ and $C_N$ appearing in Eq.~\eqref{eq:ES_xsec} of the main text in terms of the momentum transfer $Q^2$. These are given in terms of the weak neutral current form factors by~\cite{Giunti:2007ry, Formaggio:2012cpf}
\begin{equation}
\begin{split}
    A_N(Q^2) = \frac{Q^2}{m_N^2}&\,\Bigg\{\left(1+\frac{Q^2}{4m_N^2}\right)\left(G_\text{A}^{ZN}(Q^2)\right)^2 \\
    &- \left(1-\frac{Q^2}{4m_N^2}\right)\Big[\left(F_1^{ZN}(Q^2)\right)^2 \\ &~~~~~~~+ \frac{Q^2}{4m_N^2}\left(F_2^{ZN}(Q^2)\right)^2\Big]\Bigg\}
    \end{split}
    \end{equation}
    \vspace{-1em}
    \begin{equation}
    \begin{split}
        C_N(Q^2) = \frac{1}{4}\Big[&\left(G_\text{A}^{ZN}(Q^2)\right)^2 + \left(F_1^{ZN}(Q^2)\right)^2\\
        &~~~~~~~~~~+\frac{Q^2}{4m_N^2}\left(F_2^{ZN}(Q^2)\right)^2 \Big]
    \end{split}
\end{equation}
where $F_{1,2}^{ZN}(Q^2) \simeq \pm(1/2)[F_{1,2}^p(Q^2)-F_{1,2}^n(Q^2)]-2s^2_\text{W}F_{1,2}^N(Q^2)$, with 
$F_1^N$ and $F_2^N$ being respectively the Dirac and Pauli electromagnetic form factors for the nucleon $N=n,p$, with the $+\,(-)$ sign for $p\,(n)$, $s^2_\text{W} \simeq 0.229$~\cite{Workman:2022ynf} the sine squared of the Weinberg angle, $G_\text{A}^{Zp}(Q^2) \simeq  (1/2)G_\text{A}(Q^2)$ and $G_\text{A}^{Zn}(Q^2)\simeq -(1/2)G_\text{A}(Q^2)$ and $G_\text{A}(Q^2)$ the axial weak charged current form factor. We have neglected the contributions from the form factors related to strange and heavier quarks, as well as the pseudo-scalar contribution (which vanishes exactly in the case of massless neutrinos).
It is useful to define also the electric and magnetic form factors respectively as~\cite{PhysRev.119.1105, RevModPhys.35.335, Perdrisat:2006hj}
\begin{eqnarray}
G_\text{E}^N(Q^2) &\equiv& F_1^N(Q^2) - \frac{Q^2}{4m_N^2}F_2^N(Q^2),\\
G_\text{M}^N(Q^2) &\equiv& F_1^N(Q^2) +F_2^N(Q^2).
\end{eqnarray}
At zero momentum transfer, i.e.~$Q^2=0$, we have $G_\text{E}^p(0) = 1$, $G_\text{E}^n(0) = 0$, $G_\text{M}^p(0) = \mu_p/\mu_N$ and $G_\text{M}^n(0) = \mu_n/\mu_N$, where $\mu_N$ is the nuclear magneton, while $\mu_{p} \simeq 2.79\mu_N$ and $\mu_n\simeq -1.91\mu_N$ are respectively the magnetic moments of the proton and of the neutron~\cite{Workman:2022ynf}. The $Q^2$-dependence of the electric, magnetic and axial form factors is often fitted experimentally against dipole expressions, namely
$G_\text{E,M}(Q^2) = G_\text{E,M}(0)(1+Q^2/\Lambda_\text{E,M}^2)^{-2}$ with $\Lambda_\text{E,M} \simeq 0.8\,\text{GeV}$, given in terms of the experimentally measured electric charge and magnetic radii of the proton $\left\langle r_\text{E,M}^2\right\rangle^{1/2} = \sqrt{12}/\Lambda_\text{E,M} \simeq 0.85\,\text{fm}$~\cite{Alarcon:2020kcz, Gao:2021sml}, while $G_\text{A}(Q^2) = G_\text{A}(0)(1+Q^2/m_\text{A}^2)^{-2}$ with $G_\text{A}(0)\simeq 1.245$ and $m_\text{A} \simeq 1.17\,\text{GeV}$, from measurements of the axial radius $\left\langle r_\text{A}^2\right\rangle^{1/2} = \sqrt{12}/m_\text{A} \simeq 0.582\,\text{fm}$~\cite{Alexandrou:2023qbg}.

\section{Neutral Current Deep Inelastic Scattering Cross Section}\label{app:DIS}
If the centre-of-mass energy is sufficiently large, the interaction between a neutrino and a nucleon $N = p, n$ takes place with its constituents through DIS. In this section, we derive the NC DIS cross section assuming a parton model with quarks carrying a fraction $\xi$ of the nucleon's momentum, and then average the results over the quark PDFs. The process is diagrammatically represented below using the Ti\textit{k}Z-Feynamn package \cite{Ellis:2016jkw}.
\begin{center}
\begin{tabular}{cc}
\begin{tikzpicture}[baseline=(current bounding box.center)]
\begin{feynman}
\vertex(V1);
\vertex(p1)[left=2cm of V1];
\vertex(k3)[right=2cm of V1];
\node[blob, below = 1.5cm of V1](V2);
\path (V2.+180) ++ (00:-1.58) node[vertex] (p2);
\path (V2.+145) ++ (00:-1.65) node[vertex] (p2up);
\path (V2.+220) ++ (00:-1.68) node[vertex] (p2down);
\vertex(kXup)[right=1.95cm of V2];
\path (V2.-45) ++ (00:1.68) node[vertex] (kXdown);
\path (V2.+45) ++ (45:1) node[vertex] (p2q);
\path (V2.+45) ++ (45:1) node[vertex] (p2q);
\vertex[right=1cm of p2q](k4q);
\diagram*{
(p1)--[fermion, edge label=\(\nu\), pos = .6](V1),
(p2)--[fermion](V2),
(p2up)--[with arrow=0.48](V2.+145),
(p2down)--[with arrow=0.47](V2.+220),
(V1)--[boson, edge label' = \(Z\)](p2q),
(V1) --[fermion, edge label=\(\nu\), pos = .6](k3),
(V2.+45) --[fermion, edge label' = \(q\,\text{, }\bar{q}\)](p2q)--[fermion, edge label = \(q\,\text{, }\bar{q}\)](k4q),
(V2) --[fermion](kXup),
(V2.-45) --[with arrow=0.53](kXdown),
};
\vertex(X1) [right=.2cm of k4q];
\vertex(X2) [right=.2cm of kXdown];
\draw [decoration={brace}, decorate] (X1) -- (X2) node [pos=0.5, right = .2em] {\(X\)};
\vertex(N1) [left=.1cm of p2up];
\vertex(N2) [left=.1cm of p2down];
\draw [decoration={brace, mirror}, decorate] (N1) -- (N2) node [pos=0.5, left = .2em] {\(N\)};
\end{feynman}
\end{tikzpicture}
\end{tabular}
\end{center}
We fix the momenta of the incoming neutrino and quark respectively as $p_\nu =(E_\nu, \vec{p}_\nu)$ and $p_q = (E_q, \vec{p}_q)$. Analogously, for the outgoing neutrino and quarks we define $k_\nu = (E'_\nu, \vec{k}_\nu) $ and $k_q = (E'_q, \vec{k}_q
)$, respectively. Furthermore, we neglect the quark masses. The DIS is typically studied  in the frame of reference in which the nucleon is at rest $\vec{p}_N = 0$ (LAB), see, e.g., the comprehensive derivation in \cite{Giunti:2007ry} and references therein (see also \cite{Peskin:1995ev}).
However, we are interested in the frame in which the relic neutrino is at rest instead. Our approach will be that of recovering the DIS cross section in the LAB frame and write it in terms of Lorentz invariants, so to render any change of reference frame immediate. 
It will prove useful to define also the initial nucleon momentum $p_N = p_q/\xi = (E_N, \vec{p}_N)$ and the 4-momentum transfer $q = p_\nu-k_\nu$. Keeping the same notation as in the main text, we denote the squared centre-of-mass energy and momentum transfer respectively as $s = (p_N + p_\nu)^2$ and $Q^2 = -q^2 $. We also define the following Lorentz invariant quantities:
\vspace{.2em}
\begin{itemize}
\item[$\diamond$] \textit{energy transfer}
    \begin{equation} 
   \epsilon \equiv \frac{p_N\cdot q}{m_N}; 
    \end{equation}
\item[$\diamond$] \textit{inelasticity}
    \begin{equation} 
    y \equiv \frac{p_N\cdot q}{p_N\cdot p_\nu} = \frac{2 \epsilon m_N}{s-m_N^2-m_\nu^2};
    \end{equation}
\item[$\diamond$] \textit{Bjorken scaling variable}
        \begin{equation} 
    x \equiv \frac{Q^2}{2 p_N\cdot q}= \frac{Q^2}{(s-m_N^2-m_\nu^2)\, y}.
    \end{equation}
\end{itemize}
Moreover, we have the following set of relations:
\begin{eqnarray}
 Q^2 &\overset{\mathsmaller{\mathrm{LAB}}}{=}& -2m_\nu^2 + 2E_\nu E_\nu' - 2 |\vec{p}_\nu| |\vec{k}_\nu| \cos\theta,\\
 E_\nu &\overset{\mathsmaller{\mathrm{LAB}}}{=}& \frac{s-m_N^2-m_\nu^2}{2m_N},\\
 E_\nu' &\overset{\mathsmaller{\mathrm{LAB}}}{=}& E_\nu(1-y),\\
  v_\text{rel} &\overset{\mathsmaller{\mathrm{LAB}}}{=}& |\vec{p}_\nu|m_N/(E_\nu E_N),
\end{eqnarray}
where $\theta$ is the scattering angle in the LAB frame and $v_\text{rel}$ is the relative velocity between the neutrino and the nucleon. In particular, we have $\partial \cos\theta/\partial x = -m_N E_\nu y/(|\vec{p}_\nu||\vec{k}_\nu|)$ and $\partial E_\nu'/\partial y \overset{\mathsmaller{\mathrm{LAB}}}{=} -E_\nu$, so that:
\begin{equation}
    \frac{4\pi}{4E_\nu E_Nv_\text{rel}}\frac{d^3k_\nu}{(2\pi)^32E_\nu'} = \frac{y}{8\pi}\left[1+\mathcal{O}\left(\frac{m_\nu m_N}{p_\nu\cdot p_N}\right)\right]dxdy.
\end{equation}

The master formula for the DIS differential cross section can be written as \cite{Peskin:1995ev}:
\begin{widetext}
\begin{equation}\label{eq:xsec_parton}
\begin{split}
    d\sigma^{\DIS}_{\nu N} =&\,  \frac{1}{4E_\nu E_Nv_\rel}\frac{d^3k_\nu}{(2\pi)^32E_\nu'}\sum_{a = q,\bar{q}} \int_0^1 d\xi f_a^{N}(\xi, Q^2)\frac{E_N}{E_a}\int \frac{d^3k_a}{(2\pi)^32E_a'}(2\pi)^4\delta^{(4)}(\xi p_N + q - k_a)|\overline{\mathcal{M}}|^2\\
    \simeq &\,\frac{G_F^2}{2\pi}\frac{y}{(1+Q^2/M_Z^2)^2}[p_\nu^\alpha k_\nu^\beta + p_\nu^\beta k_\nu^\alpha - g^{\alpha\beta} (k_\nu\cdot p_\nu)\pm i \varepsilon^{\alpha\beta\gamma\delta}p_{\nu,\gamma}k_{\nu,\delta}]\sum_{a = q,\bar{q}} F_{\alpha\beta}^a dxdy,
    \end{split}
\end{equation}
\end{widetext}
where $+\,(-)$ applies to neutrinos (antineutrinos); $G_F\simeq 1.167\,\times 10^{-5}\text{ GeV }^{-2}$ is the Fermi weak coupling constant; $|\overline{\mathcal{M}}|^2$ is the squared Feynman amplitude of the neutrino-quark elastic scattering averaged over the initial quark spins; $g^{\alpha\beta} = \text{diag}(1, -1, -1, -1)$ is the Minkowski metric tensor; $\varepsilon^{\alpha\beta\gamma\delta}$ is the totally anti-symmetric tensor;
$f_{q(\bar{q})}^{N}$ is the Lorentz scalar PDF for the (anti)quark $q$ ($\bar{q}$) in the nucleon $N$ and we have neglected neutrino masses in comparison with the energies and other mass scales involved. The hadronic tensor $F_{\alpha\beta}^a$ related to (anti)quark $a = q (\bar{q})$ 
 is defined as
      \begin{equation}
      \begin{split}
  F^a_{\alpha\beta} = \frac{f_a^{N}(x,Q^2)}{4Q^2}\sum_\text{spins}&\,\langle a(x p_N)|J^a_\alpha|a(k_a)\rangle \\
  &\,\times
     \langle a(k_a)|J_\beta^{a\dagger}|a(x p_N)\rangle,
     \end{split}
  \end{equation}
and the neutral quark current (multiplied by the numerator of the $Z$ boson propagator and having factored out the electroweak coupling constant) as
\begin{equation}
    J_\alpha^q = \left(g_{\alpha \beta} - \frac{q_\alpha q_\beta}{M_Z^2}\right)\bar{q}\gamma^\beta (g_V^q - g_A^q\gamma_5)q,
\end{equation}
where $\gamma^\alpha$ are the ordinary Dirac gamma matrices, $\gamma^5 = (i/4!)\varepsilon_{\alpha\beta\gamma\delta}\gamma^\alpha \gamma^\beta \gamma^\gamma \gamma^\delta$ is the fifth gamma matrix, and the couplings are given by
$g_V^{u,\,c,\,t} = 1/2 - (4/3)s_{W}^2$, $   g_A^{u,\,c,\,t} = 1/2$, $g_V^{d,\,s,\,b} = -1/2 + (2/3)s_\text{W}^2$ and $    g_A^{d,\,s,\,b} = -1/2$. We note that $F^a_{\alpha \beta}$ 
can only depend on the momenta $p_N$ and $q$ and thus decomposes into
\vspace{1em}
\begin{equation}
\begin{split}
    F_{\alpha\beta}^a =& -g_{\alpha\beta}F_1^a + \frac{p_{N\,\alpha} p_{N\,\beta}}{m_N^2}\frac{m_N}{ \epsilon}F_2^a - i\,\frac{\epsilon_{\alpha\beta\gamma\delta} p^\gamma_{N} q^\delta}{2m_N^2}\frac{m_N}{\epsilon}F_3^a\\
    &+\frac{q^\alpha q^\beta}{m_N^2}F_4^a + \frac{p_{N}^\alpha q^\beta + p_{N}^\beta q^\alpha}{2m_N^2}F_5^a\\
    &+ i\, \frac{p_{N}^\alpha q^\beta - p_{N}^\beta q^\alpha}{2m_N^2} F_6^a\,.
    \end{split}
\end{equation}
With the adopted decomposition of $F_{\alpha\beta}^a$, the quantities $F^a_1$, $F^a_2$, $F^a_3$, $F^a_4$, $F^a_5$ and $F^a_6$ are dimensionless. As can be checked, the third line proportional to $F_6^a$ gives vanishing contributions when contracted with the combination of leptonic momenta appearing in the second line of Eq.~\eqref{eq:xsec_parton}. The same holds for the $F_4^a$ and $F_5^a$ contributions in the limit of vanishing neutrino masses. Thus, only the terms proportional to $F^a_1$, $F^a_2$ and $F^a_3$ are relevant in our calculation. We then get the following expressions for the $F_1^a$, $F_2^a$ and $F_3^a$ \cite{Giunti:2007ry}:
\begin{eqnarray}
    F_1^{q(\bar{q})} &=&  \frac{1}{2}[(g_V^q)^2 + (g_A^q)^2] f_{q(\bar{q})}^N(x, Q^2),\\
    F_2^{q(\bar{q})} &=& 2xF_1^{q(\bar{q})} \\
F_3^{q(\bar{q})}&=&(-)2g_V^qg_A^qf_{q(\bar{q})}^N(x, Q^2).
\end{eqnarray}
\vspace{.2em}

Plugging the above expressions in Eq.~\eqref{eq:xsec_parton}, after contraction of the Lorentz indices, we arrive to~\cite{Giunti:2007ry, Formaggio:2012cpf}:
\begin{widetext}
\begin{equation}
    \frac{d^2\sigma^{\DIS}_{\nu N}}{dxdy} \simeq
    \frac{G_F^2}{2\pi}\frac{Q^2}{ \left[1+Q^2/M_Z^2\right]^2} \Bigg[y F^{ZN}_1(x, Q^2) + \frac{1}{xy}\left(1-y-\frac{m_N^2 x^2y^2}{Q^2}\right)F^{ZN}_2(x, Q^2) 
    \pm \left(1-\frac{y}{2}\right)F^{ZN}_3(x, Q^2)\Bigg],
\end{equation}
\end{widetext}
where $+(-)$ for the scattering with (anti)neutrinos, while $F_1^{ZN} = \sum_{a=q,\bar{q}} F_1^a$, $F_2^{ZN} = \sum_{a=q,\bar{q}} F_2^a$ and $F_3^{ZN} = \sum_{a=q,\bar{q}} F_3^a$. Note that, in the above formula, $Q^2$ should be given in terms of $x$, $y$ and $s$ as $Q^2 = (s-m_N^2-m_\nu^2)xy$. It can be further simplified if we consider the sum of neutrinos and antineutrinos contributions, $\sigma^{\DIS}_\mathsmaller{(\nu+\bar{\nu})N} =\sigma^{\DIS}_{\nu N} + \sigma^{\DIS}_{\bar{\nu} N}$:
\begin{equation}
\begin{split}
    \frac{d^2\sigma^{\DIS}_\mathsmaller{(\nu+\bar{\nu})N}}{dxdy} \simeq \sum_{a=q,\bar{q}}&\,\frac{G_F^2}{2\pi}\frac{Q^2[(g_V^a)^2 + (g_A^a)^2]f_a^N(x,Q^2)}{[1+Q^2/M_Z^2]^2}\\
    &~~~~~~~~~~~~\times \left(y-2+\frac{2}{y}-\frac{2m_N^2 x^2y}{Q^2}\right).
    \end{split}
\end{equation}
In the reference frame in which the neutrino is at rest, the variable $x$ can be written as $x = (E_\nu-m_\nu)/(E_N y)$,
from which we get $dx/dE_\nu = 1/(E_N y)$.
Then, the DIS differential cross section that we need for the boosted neutrino flux calculation is given by:
\begin{equation}
    \frac{d\sigma^{\DIS}_\mathsmaller{(\nu+\bar{\nu})N}}{dE_\nu} = \frac{1}{E_N}\int_{y_\min}^{y_\max} \frac{dy}{y}\,\frac{d\sigma^{\DIS}_\mathsmaller{(\nu+\bar{\nu})N}}{dxdy},
\end{equation} 
which leads to the expression given in Eq.~\eqref{eq:DIS_xsec} of the main text,
where, to keep a shorter notation, we have written $d\sigma_{\nu N}^\DIS$ in place of $d\sigma^\DIS_\mathsmaller{(\nu+\bar{\nu})N}$.
As already specified, to evaluate numerically the quark PDFs, we made use of the Python package \texttt{parton} and the \virg{CT10} PDF set~\cite{Lai:2010vv}.
The quark PDFs in the considered library are given for $x\geq x_\min = 10^{-8}$ and in the range $Q^2_\min = 1.69\,\text{GeV}^2\leq Q^2 \leq Q^2_\max= 10^{10}\,\text{GeV}^2$. This practically limits our ability to properly estimate the DIS cross section outside the energy range $Q^2_\min/(2m_\nu)\leq E_\nu -m_\nu\leq \min\{Q^2_\max/(2m_\nu), T_\nu^\max(T_N)\}$, where $T_\nu^\max(T_N)$ is the maximal kinetic energy the neutrino can have after an elastic scattering,
\begin{equation}\label{eq:Tnumax}
T_\nu^\max(T_N) = \frac{\left(T_N^2 + 2m_N T_N\right)}{T_N + (m_N+m_\nu)^2/(2m_\nu)},
\end{equation}
and $T_N\equiv E_N-m_N$ the initial kinetic energy of the incoming nucleon.
As an indicative procedure to avoid
unphysical discontinuities, we have extrapolated the DIS contribution for $Q^2 \leq Q^2_{\min}$ by assuming a linear scaling
in $Q^2$. In practice, this does not affect any of our predictions for the range of $E_\nu$ of interest for the UHE neutrino telescopes considered in our study. Finally, we note that the inelasticity parameter lies between $y_{\min} =(E_\nu-m_\nu)/E_N$ and $y_{\max} = \min\{1, (E_\nu-m_\nu)/(E_Nx_\min)\} = 1$. 

\section{Boosted Relic Neutrinos from Cosmic Rays: Revisited}\label{app:comparison}
We revisit the computation of the CR-induced boosted relic neutrino flux discussed in~\cite{Ciscar-Monsalvatje:2024tvm}, implementing the full cross section with $Q^2$-dependence of the form factors, the DIS contribution and a mixed CR composition. Firstly, we note that the ES cross section in the low-energy limit $2m_\nu E_N \lesssim m_N^2$~\cite{Giunti:2007ry} reads:
\begin{equation}\label{eq:xsec_lowEn}
    \frac{d\sigma_{\nu N}^\ES}{dE_\nu}\simeq \frac{G_F^2(s-m_N^2)^2}{16\pi s\,T_\nu^\max(T_N)}[(1-4s_\text{W}^2)^2+3\,(G_\text{A}(0))^2],
\end{equation}
with $T_\nu^\max(T_N)$ as given in Eq.~\eqref{eq:Tnumax}.
The expression in Eq.~\eqref{eq:xsec_lowEn} resembles the expression reported in~\cite{Ciscar-Monsalvatje:2024tvm} in the same limit, apart from an overall $\mathcal{O}(1)$ factor. Our calculation of the cross section differs instead more with respect to the one in~\cite{Ciscar-Monsalvatje:2024tvm} for  $2m_\nu E_N \gtrsim m_N^2$.

\begin{figure}
    \centering
    \includegraphics[width = .48\textwidth]{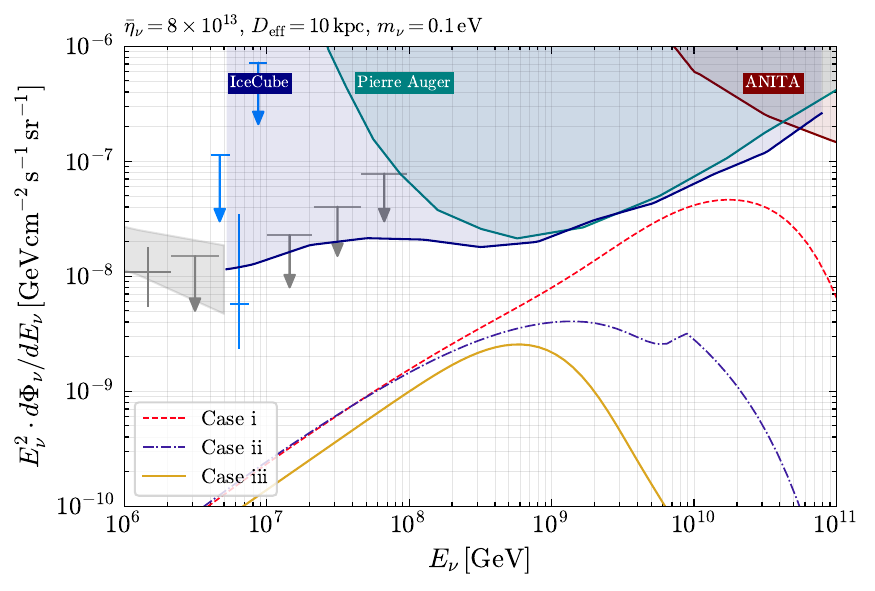}
    \caption{Comparison between different procedures used to calculate the boosted all-flavour relic neutrino flux according to Eq.~\eqref{eq:dPhinudE} of the main text, with $\bar{\eta}_\nu = 8\,\times 10^{13}$, $D_\text{eff}=10\,\text{kpc}$ and $m_\nu = 0.1\,\text{eV}$. The dashed red line corresponds to case i) for which a low-energy limit of the elastic scattering cross section and a pure proton CR flux are considered; the dot-dashed purple line to case ii) with a full elastic plus inelastic cross section and a pure proton CR flux; the thick yellow line to case iii) with a full cross section and a mixed CR composition according to Scenario 1 of~\cite{PierreAuger:2022atd}. See the text for further details. The excluded regions and data points are as in Fig.~\ref{fig:Boosted_nu_flux}.}
    \label{fig:fluxes_comparison}
\end{figure}

We show in Fig.~\ref{fig:fluxes_comparison} the neutrino flux computed according to Eq.~\eqref{eq:dPhinudE}, with $m_\nu = 0.1\,\text{eV}$, $D_\text{eff}=10\,\text{kpc}$, $\bar{\eta}_\nu = 8\,\times 10^{13}$ roughly saturating the limit given in~\cite{Ciscar-Monsalvatje:2024tvm}, and different benchmark cases described in what follows. Our choice of $D_\text{eff}$ is motivated if the neutrino overdensity is localised on a scale of the order of the MW radius.
\begin{itemize}
\item[i)] The red dashed line is obtained by taking the cross section as in \eqref{eq:xsec_lowEn} after recasting the all-species CR flux as observed at Earth from~\cite{Lv:2024wrs} in the energy range $10^5\,\text{GeV}\lesssim E_\CR\lesssim 200\,\text{EeV}$ and assuming a pure proton composition from the lowest to the largest energies. With this procedure we obtain results that are comparable (but not overlapping) with those presented in~\cite{Ciscar-Monsalvatje:2024tvm}.
\item[ii)] The dot-dashed purple line is obtained by considering the full $Q^2$-dependent cross section as a sum of the elastic and deep inelastic contributions, as discussed in the main text, while keeping the pure proton CR flux as in case i). We note a suppression compared to case i) starting from $E_\nu\sim\text{EeV}$ due to the form factors $A_N(Q^2)$ and $C_N(Q^2)$ of the elastic cross section, as well as a kick at $E_\nu \sim 10\,\text{EeV}$ because of the DIS.
\item[iii)] The thick yellow curve is obtained by considering the full cross section as in case ii) and a mixed composition for the CR flux according to the analysis presented in~\cite{PierreAuger:2022atd}. Concentrating on the extragalactic contribution only, we considered this flux in the energy range $10^8\,\text{GeV}\lesssim E_\CR \lesssim 200\,\text{EeV}$. For definiteness, we focused on the Scenario 1 described in~\cite{PierreAuger:2022atd}, in which the proton composition gets suppressed earlier compared to heavier nuclei, thus implying an attenuation of the boosted relic neutrino flux.
\end{itemize}
We point out that the analogous flux in Fig.~\ref{fig:Boosted_nu_flux} is obtained as in case iii), but considering three neutrino species with masses that saturate the cosmological bound \cite{DESI:2024mwx} -- while that in Fig.~\ref{fig:fluxes_comparison} is for three degenerate neutrinos with reference mass $m_\nu = 0.1\,\text{eV}$ -- and starting the integration from $E_\CR^\text{ankle}$ to remain conservative on the galactic-to-extragalactic transition.

 We find that the implementation of the full momentum transfer dependence of the neutrino-nucleus cross section, the deep inelastic scattering contribution and a mixed CR composition, leads to a boosted relic neutrino
flux suppressed significantly with respect to what
reported in \cite{Ciscar-Monsalvatje:2024tvm}, and, correspondingly, the bounds on the
local neutrino overdensity are weakened by an overall
$\mathcal{O}(10)$ factor.

\section{Normalisation of the Cosmic Ray Flux 
Inside Galaxy Clusters}\label{app:normalisation}
We discuss here the normalisation procedure that we followed for our estimate of the CR flux within galaxy clusters acting as CR-reservoirs, i.e.~to determine the function $d\Phi_{\CR}/dE_{\CR}$ entering Eq.~\eqref{eq:dPhinudE} of the main text.
We remind that Eq.~\eqref{eq:dPhinudE} is intended to be a sum over all species $\mathcal{N}$ constituting CRs, where the dependence on $\mathcal{N}$ is contained both in the flux and in the differential cross section. Concerning the flux, one could proceed by summing
the contribution from all CR-reservoirs in the entire Universe, but this requires knowledge on details of the population of clusters that is not really necessary if one assumes that most UHECR observed on Earth originate in these CR-reservoirs. Therefore, under this assumption
and as we explain below, we extract the flux from \cite{Fang:2017zjf}, where it has been determined by comparison with the observed UHECR.

We rewrite here the flux of cosmic nucleus $\mathcal{N}$ as given in~\cite{Fang:2017zjf} and as used in our main analysis:
\begin{equation}
\frac{d\Phi_{\mathcal{N}}}{dE_\mathcal{N}} = K_\mathcal{N}\left(\frac{R_\max}{R}\right)^\alpha e^{-R/R_\max}.
\end{equation}
where the rigidity parameter $R$ is defined as $R \equiv E_\mathcal{N}/(Z_\mathcal{N}q_e)$, $q_e$ being the electric charge unit, with $R_\max = 2\times 10^{21}/26\,\text{eV}$~\cite{Fang:2017zjf} (for convenience, we absorb the electric charge $q_e$ in the definition of $R$ so that it is measured in eV rather than V). According to~\cite{Fang:2017zjf}, at fixed $R$ the chemical composition of $E_\mathcal{N}^2d\Phi_{\mathcal{N}}/dE_\mathcal{N}$ is (0.625, 0.252, 0.053, 0.009, 0.124) for (${}^1\text{H}$, ${}^4\text{He}$, CNO, ${}^{28}\text{Si}$, ${}^{56}\text{Fe}$) respectively (as reference for CNO we have considered ${}^{14}\text{N}$ only). Thus, at fixed $R$, we have
\begin{equation}
E_\mathcal{N}^2\frac{d\Phi_\mathcal{N}}{dE_\mathcal{N}}= Z_{\mathcal{N}}^2K_\mathcal{N}R^2\left(\frac{R_\max}{R}\right)^{\alpha}e^{-R/R_{\text{max}}},
\end{equation}
so that the aforementioned proportions are reflected in the ratios of $Z_{\mathcal{N}}^2K_\mathcal{N}$. Therefore, $K_\mathcal{N}\equiv  (C_\text{N}/Z_\text{N}^2)K_\text{H}$, with  $C_\text{He} \simeq 0.4032$, $C_\text{N} \simeq0.0848$, $C_\text{Si} \simeq0.0144$ and $C_\text{Fe} \simeq 0.1984$. The only parameter left to determine is $K_H$, which we fix by requiring that the luminosity emitted from the population of CR-reservoirs matches with the one observed at Earth, by, e.g., Pierre Auger above the ankle. More specifically, we impose
\begin{equation}\label{eq:fix_norm}
\sum_{\mathcal{N}}\int_{E^\text{ankle}_\CR}^{E_\CR^\max} dE_\mathcal{N} E_\mathcal{N}\frac{d\Phi_\mathcal{N}}{dE_\mathcal{N}} = \int_{E^\text{ankle}_\CR}^{E_\CR^\max} dE_\CR E_\CR \frac{d\Phi_\CR}{dE_\CR},
\end{equation}
where the CR flux at Earth is taken from~\cite{PierreAuger:2020kuy} to be $d\Phi_\CR/dE_\CR \simeq (d\Phi_\CR/dE_\CR)_\text{ankle}\, (E_\CR^\text{ankle}/E_\CR)^{2.5}$, with $(d\Phi/dE_\CR)_\text{ankle} \simeq 10^{-27}\,\text{GeV}^{-1}\text{cm}^{-2}\text{s}^{-1}\text{sr}^{-1}$, within the range $E^\text{ankle}_\CR = 5\,\text{EeV}\leq E_\CR\leq E_\CR^\max = 200\,\text{EeV}$. The CR flux observed at Pierre Auger and reported in~\cite{PierreAuger:2020kuy} is actually more complicated than the single power-law considered above, including changes in the slopes at and above the ankle, mixed composition and a suppression at $\sim 50\,\text{EeV}$. We have checked, however, that such additional features do not change appreciably the value of the integral in the right-hand-side of Eq.~\eqref{eq:fix_norm}, to which the shape at lower energies dominates. The integral on the left-hand-side can be computed analytically as follows:

\begin{widetext}
\begin{eqnarray}
\nonumber\sum_{\mathcal{N}} \int_{E^\text{ankle}_\CR}^{E_\CR^\max} dE_\mathcal{N} E_\mathcal{N}\frac{d\Phi_\mathcal{N}}{dE_\mathcal{N}} &=& K_{\text{H}}\sum_{\mathcal{N}} \frac{C_\mathcal{N}}{Z_{\mathcal{N}}^2}\int_{E_\CR^\text{ankle}}^{E^\max_\CR}dE_\mathcal{N}\; E_\mathcal{N} \left(\frac{E^\max_\mathcal{N}}{E_\mathcal{N}}\right)^{\alpha}e^{-E_\mathcal{N}/E^\max_\mathcal{N}} \\
\nonumber& =& K_{\text{H}}\sum_{\mathcal{N}} C_\mathcal{N}R_{\text{max}}^2\int_{E_\CR^\text{ankle}/(Z_\mathcal{N}R_\max)}^{E^\max_\CR/(Z_\mathcal{N}R_\max)}dx\; x^{1-\alpha}e^{-x}\\
&=& K_\text{H} \sum_\mathcal{N} C_\mathcal{N} R_\max^2 \left[\Gamma\left(2-\alpha,\frac{E^\text{ankle}_\CR}{Z_\mathcal{N}R_\text{max}}\right) - \Gamma\left(2-\alpha,\frac{E^\max_\CR}{Z_\mathcal{N}R_\text{max}}\right)\right], 
\end{eqnarray}
\end{widetext}
where $\Gamma(a,z)$ is the upper incomplete Gamma function. For different values of the slope $\alpha$, we find the normalisation factors listed in Table \ref{tab:K_values}. These normalisation
factors \textit{de facto} encode all the information about CR-reservoirs
(distance, spatial distribution, CR-spectrum
escaping them), that is not needed for our purposes as long as one assumes (as done in \cite{Fang:2017zjf}) that most UHECRs observed on Earth originate from CR-reservoirs.

\newcolumntype{C}{@{}>{\centering\arraybackslash}X}
\begin{table}[h!]
\centering
\begin{tabularx}{\linewidth}{@{}|C|C|C|C|@{}}
        \hline
        \rowcolor[gray]{.95}
    \multicolumn{4}{|@{}c@{}|}{\bf Normalisation factors}\\
    \hline
    \hline
    \rule{0pt}{1em} 
        &
        $\alpha = 2$ & $\alpha = 2.3$ & $\alpha = 2.5$ \\
        \hline
    \rule{0pt}{1em} 
        $K_\text{H}$
        & $1.94\times 10^{-30}$ &$9.44\times 10^{-31}$ & $5.20\times 10^{-31}$ \\
        $K_\text{He}$
        & $1.96\times 10^{-31}$&  $9.52\times 10^{-32}$ & $5.24\times 10^{-32}$ \\
        $K_\text{N}$
        & $3.36 \times 10^{-33}$ & $1.63 \times 10^{-33}$ & $8.99\times 10^{-34}$ \\
        $K_\text{Si}$
        & $1.43 \times 10^{-34}$ & $6.94 \times 10^{-35}$ & $3.82\times 10^{-35}$ \\
        $K_\text{Fe}$
        &$5.70\times 10^{-34}$ & $2.77\times 10^{-34}$ & $1.52\times 10^{-34}$ \\
        \hline
    \end{tabularx}
    \caption{Flux normalisation factors for each nuclear species considered and different benchmark values of $\alpha$. All listed values are in units of $\text{GeV}^{-1}\text{cm}^{-2}\text{s}^{-1}\text{sr}^{-1}$.}
    \label{tab:K_values}
\end{table}

\end{document}